\documentclass[12pt,a4paper,twoside]{article}

\usepackage[utf8]{inputenc}
\usepackage{graphicx}

\title{Charge effects}
\author{Yahya Wasiu Akanni}
\date{4th April 2014}

\usepackage{color}
\usepackage{amsfonts}
\usepackage{lscape}
\usepackage{hyperref}
\usepackage{epstopdf}
\usepackage{lscape}
\usepackage{amsmath}
\usepackage{hyperref}
\usepackage{amsmath}
\usepackage{setspace}
\usepackage{cite}
\begin{document}
\begin{center}
\textbf{Position and momentum information-theoretic measures of the pseudoharmonic potential}
\end{center}
\vspace{4mm}

\begin{center}

Yahya, W. A. $^\dagger$ $^\ddagger$\footnote{wazzy4real@yahoo.com}, Oyewumi, K. J. $^\dagger$\footnote{kjoyewumi66@unilorin.edu.ng}, Sen, K. D. $^\diamond$ \footnote{sensc@uohyd.ernet.in}.\\

\vspace{5mm}
 $^\dagger$ { Theoretical Physics Section, Department of Physics, University of Ilorin, Nigeria. \\}

$^\ddagger$ { Department of Physics and Material Science, Kwara
State University, Malete, Nigeria. \\}

$^\diamond$ { School of Chemistry, University of Hyderabad, Hyderabad 500046, India \\} 

\end{center}

\begin{abstract}
In this study, the information-theoretic measures in both the position and momentum spaces for the pseudoharmonic potential using Fisher information, Shannon entropy, Renyi entropy, Tsallis entropy and Onicescu information energy  are investigated analytically and numerically. The results obtained are applied to some diatomic molecules. The Renyi and Tsallis entropies are analytically obtained in position space using Srivastava-Niukkanen linearization formula in terms of the Lauricella hypergeometric function. Also they are obtained in the momentum space in terms of the multivariate Bell polynomials of Combinatorics. We observed that the Fisher information increases with $n$ in both the position and momentum spaces, but decreases with $\ell$ for all the diatomic molecules considered. The Shannon entropy also increases with increasing $n$ in the position space and decreases with increasing $\ell$. The variations of the Renyi and Tsallis entropies with $\ell$ are also discussed. The exact and numerical values of the Onicescu information energy are also obtained, after which the ratio of information-theoretic impetuses to lengths for Fisher, Shannon and Renyi are obtained.
\end{abstract}

\section{Introduction}

Information-theoretic methods have been used extensively to study various systems in communications \cite{Sha48}, Physics \cite{Bia75}, Chemistry \cite{Sea80,Kar04}, Biology \cite{Ada04,Pan08}. The various information-theoretic quantities such as Shannon information entropy, Fisher information, Renyi entropy, Tsallis entropy, Onicescu information energy have been discussed for different quantum mechanical systems \cite{Deh01, DeL05, Rom08, Lop08, Fra10, Apt10, Pat07, Pat07a, Pat07c, Pat12, San11}. An interesting and simple method has been used to obtain Shannon, Fisher, Onicescu and Tsallis entropies in atoms as functions of the atomic number Z in ref. \cite{Pan08}. 

Fractional occupation probabilities of electrons in atomic orbitals were employed, instead of the more complicated continuous electron probability densities. The information-theoretic quantities have also been discussed by using scaling properties of the systems \cite{Pat07, Pat07a, Pat07c, Pat12, Gon09, Sza08}. The relativistic effect have also been considered in the study of the information-theoretic quantities \cite{Jac10, Man10, Mal10, MaD10, Bor07, San09,  Yah13b}. The Shannon information entropy $S[\rho]$ of the electron density, $\rho(\textbf{r})$, in the coordinate space is defined as \cite{Sen06, Deh98}
\begin{equation}
S[\rho]=-\int \rho(\textbf{r}) \ln \rho(\textbf{r}) d\textbf{r},
\label{ps1}
\end{equation}
and the corresponding momentum space Shannon entropy $S[\gamma]$ is given by
\begin{equation}
S[\gamma]=-\int \gamma(\textbf{p}) \ln \gamma(\textbf{p}) d\textbf{p},
\label{ps2}
\end{equation}
where $\gamma(\textbf{p})$ denotes the momentum density. The Shannon entropy measures the uncertainty in the localization of a particle in space \cite{Deh01}. 

The Fisher information entropy is defined in the position space as \cite{DeL05}
\begin{equation}
I[\rho]=\int \frac{\left[ \nabla \rho(\textbf{r})\right] ^{2}}{\rho (\textbf{r})}d\textbf{r},
\label{ps3}
\end{equation}
and the corresponding momentum space measure is given by \cite{Jac10}
\begin{equation}
I[\gamma]=\int \frac{\left[ \nabla \gamma(\textbf{p})\right] ^{2}}{\gamma (\textbf{p})}d\textbf{p}.
\label{ps4}
\end{equation}
The Fisher information entropy in the position space measures the narrowness and the oscillation nature of the probability distribution \cite{Mon08}. 

The Onicescu information energy is defined in the position and momentum spaces, respectively, as \cite{Cha05}
\begin{equation}
E[\rho]=\int \rho^{2}(\textbf{r})d\textbf{r},
\label{ps5}
\end{equation}
and
\begin{equation}
E[\gamma]=\int \gamma^{2}(\textbf{p})d\textbf{p}.
\label{ps6}
\end{equation}
The greater the information energy, the more concentrated is the probability distribution, while the information content decreases \cite{Cha05}. The Renyi position entropy defined as \cite{Rom08, Lop08, Fra10, Apt10}
\begin{equation}
R_{q}[\rho]=\frac{1}{1-q}\log \left[ \int \rho(\textbf{r})^{q}d\textbf{r}\right] \; 0<q<\infty, \; q\neq 1,
\label{ps7}
\end{equation}
is a generalization of the Shannon entropy. The Renyi entropy in the momentum space is also given as
\begin{equation}
R_{q}[\gamma]=\frac{1}{1-q}\log \left[ \int \gamma(\textbf{p})^{q}d\textbf{p}\right] \; 0<q<\infty, \; q\neq 1.
\label{ps8}
\end{equation}
The Tsallis entropy is defined in the position and momentum spaces, respectively, as \cite{Pat07, Pat12, Gon09, Por04, Grau10, Rob97}
\begin{equation}
T_{q}[\rho]=\frac{1}{q-1}\left(1-\int \rho(\textbf{r})^{q}d\textbf{r} \right)
\label{ps9}
\end{equation}
and 
\begin{equation}
T_{m}[\gamma]=\frac{1}{m-1}\left(1-\int \gamma(\textbf{p})^{m}d\textbf{p} \right),\; \frac{1}{q}+\frac{1}{m}=2.
\label{ps10}
\end{equation}

The pseudoharmonic potential has been very useful in the area of physical sciences and it has been extensively used to describe interaction of some diatomic molecules since its introduction \cite{Oye12, Don07, Oye08, Sag85, Oye03, Ikh07}. This potential is used to describe the roto-vibrational states of diatomic molecules and nuclear rotations and vibrations \cite{Pat07a}. The pseudoharmonic potential can be written as \cite{Oye12}
\begin{equation}
V(r)=D_{e}\left(\frac{r}{r_{e}}-\frac{r_{e}}{r} \right)^{2}, 
\label{ps11}
\end{equation}
where $D_{e}$ is the dissociation energy between two atoms in a solid and $r_{e}$ is the equilibrium inter-molecular separation. The radial part of the Schr\"{o}dinger equation for the pseudoharmonic potential is written as ($\hbar=\mu=1$) \cite{Oye12}:
\begin{equation}
\left\lbrace \frac{d^{2}}{dr^{2}}+\frac{2}{r}\frac{d}{dr}+\left[2E-2\left(\frac{D_{e}r^{2}}{r_{e}^{2}}+\frac{D_{e}r_{e}^{2}}{r^{2}}-2D_{e} \right) -\frac{\ell(\ell+1)}{r^{2}}  \right]  \right\rbrace R_{n,\ell}(r) =0.
\label{ps12}
\end{equation}

By solving equation (\ref{ps12}), Oyewumi and Sen obtained the wave function as \cite{Oye12}
\begin{equation}
R_{n,\ell}(r)=N_{n,\ell} \,r^{\gamma_{_{\ell}}}e^{-\lambda r^{2}}L_{n}^{\gamma_{_{\ell}}+\frac{1}{2}}(2\lambda r^{2}),
\label{ps13}
\end{equation}
where
\begin{equation}
 \gamma_{_{\ell}}=\frac{1}{2}\left[-1+\sqrt{(2\ell+1)^{2}+8 D_{e}r_{e}^{2}} \right],~~
 \lambda=\sqrt{\frac{D_{e}}{2r_{e}^{2}}},~~
 N_{n,\ell}=\left[\frac{2(2\lambda^{2})^{\frac{1}{4}(2\gamma_{_{\ell}}+3)}n!}{\Gamma\left(n+\gamma_{_{\ell}}+\frac{3}{2} \right) } \right]^{\frac{1}{2}}. 
 \label{ps14}
 \end{equation}
The wave function in equation (\ref{ps13}) is the position space wave function. The momentum space wave function is obtained as \cite{Yan94}
\begin{equation}
R_{n,\ell}(p)= \left[\frac{2n! (2 \lambda^{2})^{\frac{-\gamma_{_{\ell}}}{2}-\frac{3}{4}}}{\Gamma\left( n+\gamma_{_{\ell}}+\frac{3}{2}\right) } \right] ^{\frac{1}{2}} p^{\gamma_{_{\ell}}}e^{-\frac{p^{2}}{\lambda}}L_{n}^{\gamma_{_{\ell}}+\frac{1}{2}}\left( \frac{p^{2}}{2 \lambda} \right) .
\label{ps15}
\end{equation}
The probability densities in the position space and the momentum space are therefore written, respectively, as
\begin{equation}
\rho(r)= N^{2}_{n,\ell}  \,r^{2 \gamma_{_{\ell}}} \, e^{-2 \lambda r^{2}}\left[ L_{n}^{\gamma_{_{\ell}}+\frac{1}{2}}(2\lambda r^{2}) \right] ^{2}
\label{ps16}
\end{equation}
and
\begin{equation}
\gamma(p)= \frac{2n! (2 \lambda)^{-\gamma_{_{\ell}}-\frac{3}{2}}}{\Gamma\left( n+\gamma_{_{\ell}}+\frac{3}{2}\right) } \, p^{2 \gamma_{_{\ell}}}e^{-2 \frac{p^{2}}{\lambda}} \left[ L_{n}^{\gamma_{_{\ell}}+\frac{1}{2}}\left( \frac{p^{2}}{2 \lambda} \right)\right] ^{2}.
\label{ps17}
\end{equation}

In this paper, we shall study the pseudoharmonic potential using the information-theoretic measures defined above. The paper is organised as follows. In Section \ref{sec2}, the Fisher and Shannon information entropies for the pseudoharmonic potential are obtained. Section \ref{sec3} contains the Renyi entropy, Tsallis entropy and Onicescu information energy for the pseudoharmonic potential. The information theoretic lengths and impetuses are discussed in Section \ref{sec4} while the Conclusion is given in Section \ref{sec5}. 

\section{Fisher and Shannon Information Entropies for the pseudoharmonic Potential}
\label{sec2}
\subsection{Fisher information entropy for the pseudoharmonic potential}
From equation (\ref{ps3}), the Fisher information entropy in the position space can be written as
\begin{equation}
I[\rho]=4\pi \int_{0}^{\infty}\frac{r^{2}}{\rho(r)}\left(\frac{\partial \rho}{\partial r} \right)^{2}dr, 
\label{ps18}
\end{equation}
where $\rho(r)$ is given in equation (\ref{ps16}). The integral in equation (\ref{ps18}) is obtained, after some manipulations, as
\begin{equation}
I[\rho]=\frac{2^{ (13 - 2 \gamma_{\ell})/4 }\pi \lambda  }{1+2\gamma_{_{\ell}}}\left[ (4n+3)+(8n+4)\gamma_{_{\ell}}\right]. 
\label{ps19}
\end{equation}
In the momentum space, the Fisher information is
\begin{equation}
I[\gamma]=4\pi \int_{0}^{\infty}\frac{p^{2}}{\gamma(p)}\left(\frac{\partial \gamma}{\partial p} \right)^{2}dp,
\label{ps20}
\end{equation}
where $\gamma(p)$ is as given in equation (\ref{ps17}). For ground state ($n=0, \ell=0$), equation (\ref{ps20}) is obtained as
\begin{equation}
I[\gamma]=\frac{2^{(7 - 6\gamma_{\ell})/4   }\pi \left( 3+4 \gamma_{_{\ell}}\right)} {\lambda\left(1+2\gamma_{_{\ell}}\right) }.
\label{ps21}
\end{equation}

The Fisher information for the pseudoharmonic potential is obtained for some diatomic molecules, the results are given in Tables \ref{tab2} and \ref{tab3}. The values of the dissociation energy $D_{e}$ and equilibrium intermolecular separation $r_{e}$ of the diatomic molecules used in this work are given in Table \ref{tab1},  these are obtained from Refs. \cite{Oye12, Ikh07, Sev07, Ber09, Har00, Jia13}.
It can be observed from Tables \ref{tab2} and \ref{tab3} that the Fisher information (both in the position and momentum spaces) increase with $n$ for the diatomic molecules considered. 

\begin{figure}[h]
\caption{Variation of the Fisher information entropy in the position space $(I[\rho])$ with $\ell$ for $Na_{2}$, $Cl_{2}$, $O_{2}^{+}$, $N_{2}^{+}$, and $NO^{+}$. }
\centering
\includegraphics[scale=0.7]{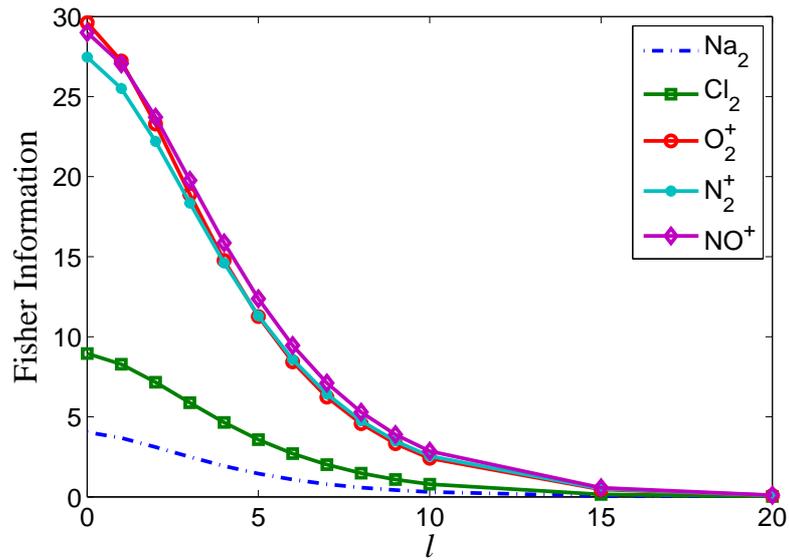}
\label{fig1}
\end{figure}

The variations of the Fisher information with orbital angular momentum $\ell$ are shown in Figure \ref{fig1} with $n=0$ for five selected diatomic molecules. It is observed from Figure \ref{fig1} that the Fisher information in the position space decreases with increasing $\ell$. The values of the Fisher information entropy for all the diatomic molecules considered tend to $0$ for $\ell\gg 1$.

\subsection{The Shannon Information Entropy of the pseudoharmonic Potential}
\label{sec2.2}

\begin{figure}[h]
\caption{Variation of the Shannon information entropy $S[\rho]$ with $\ell$ for $Na_{2}$, $Cl_{2}$, $O_{2}^{+}$, $N_{2}^{+}$, and $NO^{+}$.}
\centering
\includegraphics[scale=0.7]{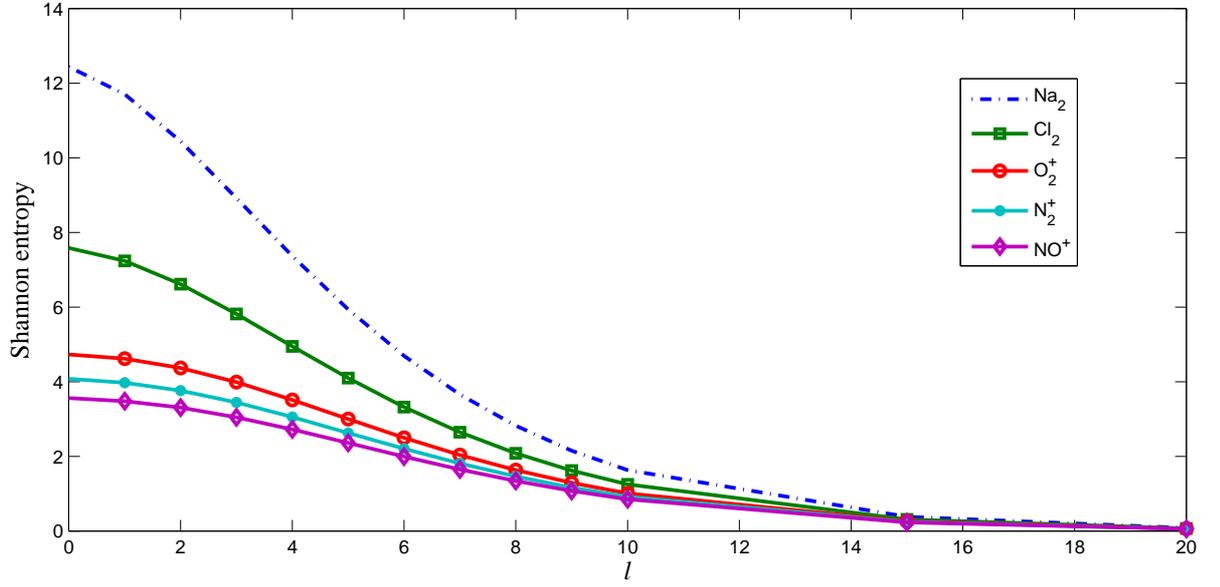}
\label{fig2}
\end{figure}

By using equation (\ref{ps16}), the position space Shannon information entropy
\begin{equation}
S[\rho]=-4 \pi \int_{0}^{\infty} r^{2} \rho(r) \ln \rho(r) dr
\label{ps22},
\end{equation}
is obtained as \cite{Deh01, Yan94}
\begin{equation}
\begin{array}{c}
 S_{n,\ell}[\rho]=2n+\gamma_{_{\ell}}+\frac{3}{2}-\ln\left(\frac{2n!}{\Gamma\left(n+\gamma_{_{\ell}}+\frac{3}{2} \right) } \right)-\gamma_{_{\ell}}\psi\left(n+\gamma_{_{\ell}}+\frac{3}{2} \right) \\ 
-\frac{3}{2}\ln(2 \lambda)+E\left( \tilde{L}_{n}^{\gamma_{_{\ell}}+\frac{1}{2}} \right),
 \end{array}  
\label{ps23}
\end{equation}
where 
\begin{equation}
E\left( \tilde{L}_{n}^{\gamma_{_{\ell}}+\frac{1}{2}} \right)= \int_{0}^{\infty}t^{\gamma_{_{\ell}}+\frac{1}{2}}e^{-t}\left[\tilde{L}_{n}^{\gamma_{_{\ell}}+\frac{1}{2}}(t) \right]^{2} \ln\left[ \tilde{L}_{n}^{\gamma_{_{\ell}}+\frac{1}{2}}(t) \right]^{2}dt
\label{ps24}
\end{equation}
is the entropic integral of the orthonormal Laguerre polynomials, which can be obtained in one dimension as \cite{Deh01}
\begin{equation}
E\left( \tilde{L}_{n}^{\gamma_{_{\ell}}+\frac{1}{2}} \right)= -2n+\left(\gamma_{_{\ell}}+\frac{3}{2} \right)\ln n-\left(\gamma_{_{\ell}}+\frac{1}{2} \right)-2+\ln(2 \pi)+o(1).   
\label{ps25} 
\end{equation}
The momentum space Shannon information entropy
\begin{equation}
S[\gamma]=-4 \pi \int_{0}^{\infty} r^{2} \gamma(p) \ln \gamma(p) dp,
\label{ps26}
\end{equation}
is obtained as \cite{Deh01, Yan94}
\begin{equation}
S_{n,\ell}[\gamma]=2n+\gamma_{_{\ell}}+\frac{3}{2}-\ln\left(\frac{2n!}{\Gamma\left(n+\gamma_{_{\ell}}+\frac{3}{2} \right) } \right)-\gamma_{_{\ell}}\psi\left(n+\gamma_{_{\ell}}+\frac{3}{2} \right)+\frac{3}{2}\ln(2 \lambda)+E\left( \tilde{L}_{n}^{\gamma_{_{\ell}}+\frac{1}{2}} \right).
\label{ps27}
\end{equation}
Table \ref{tab4} contains the numerical values of the Shannon entropy in the position space by making use of equation (\ref{ps22}). We observe that the Shannon entropy in the position space increases with $n$. The numerical values of the Shannon entropy in the momentum space are shown in Table \ref{tab5}.
The variations of the Shannon entropy with $\ell$ are shown in the Figure \ref{fig2} for the diatomic molecules considered. We observe that the Shannon entropy $S[\rho]$ decreases with $\ell$ and tend to zero for large $\ell$.

\section{Renyi entropy, Tsallis entropy and Onicescu Information energy for the pseudoharmonic potential}
\label{sec3}

In this Section, we shall calculate the Renyi entropy, Tsallis entropy and Onicescu information energy for the pseudoharmonic potential in the position and momentum spaces.

\subsection{Renyi Entropy in position space}

The Renyi entropy for the pseudoharmonic potential is written in the position space as
\begin{equation}
R_{q}[\rho]=\frac{1}{1-q}\log W_{q}[\rho],
\label{ps28}
\end{equation}
where
\begin{equation}
W_{q}[\rho]=4 \pi \, \int _{0}^{\infty} r^{2} \rho(r)^{q}dr.
\label{ps29}
\end{equation}
If we let $2 \lambda r^{2}=x$, then, $W_{q}[\rho]$ can be written as
\begin{equation}
W_{q}[\rho]=\frac{2 \pi N_{n,\ell}^{2q}}{(2 \lambda)^{q \gamma_{_{\ell}}+\frac{3}{2}}} \int_{0}^{\infty}x^{q \gamma_{_{\ell}}+\frac{1}{2}} \, e^{-qx} \left[L_{n}^{\gamma_{_{\ell}}+\frac{1}{2}}(x) \right]^{2q}dx=\frac{2 \pi N_{n,\ell}^{2q}}{(2 \lambda)^{q \gamma_{_{\ell}}+\frac{3}{2}}} F_{q}[\rho].
\label{ps30} 
\end{equation}

To calculate the integral $F_{q}[\rho]=\int_{0}^{\infty}x^{q \gamma_{_{\ell}}+\frac{1}{2}} \, e^{-qx} \left[L_{n}^{\gamma_{_{\ell}}+\frac{1}{2}}(x) \right]^{2q}dx$, we make use of the linearization formula of Srivastava-Niukkanen for the products of various Laguerre polynomials given by \cite{San11, Sri88, Sri03}
\begin{equation}
 x^{\mu}L_{m_{1}}^{(\alpha_{1})}(t_{1}x).....L_{m_{r}}^{(\alpha_{r})}(t_{r} x)=\sum_{k=0}^{\infty}\Theta_{k}\left(\mu,\beta, r, {m_{i}}, {\alpha_{i}},{t_{i}} \right) L_{k}^{(\beta)}(x),
 \label{ps31}
 \end{equation} 
where
\begin{equation}
\Theta_{k}\left(\mu,\beta, r, {m_{i}}, {\alpha_{i}},{t_{i}} \right)=\begin{array}{l}
(\beta+1)_{\mu}\left(\begin{array}{c}
\alpha_{1}+m_{1} \\ 
m_{1}
\end{array}  \right)...
\left(\begin{array}{c}
\alpha_{n}+m_{n} \\ 
m_{n}
\end{array}  \right)\times  \\ 
F_{A}^{(r+1)} \left[ \beta+\mu+1,-m_{1},...,-m_{n},-k;\alpha_{1}+1,...,\alpha_{n}+1,\beta+1;t_{1},...,t_{n},1 \right],
\end{array}
 \label{ps32}
\end{equation}
in terms of Lauricella's hypergeometric functions of $(r+1)$ variables. The Pochhammer symbol $(a)_{n}=\frac{\Gamma(a+n}{\Gamma(a)}$ and the binomial number $\left(\begin{array}{c}
a\\
b
\end{array} \right) =\frac{\Gamma(a+1)}{\Gamma(b+1)}\Gamma(a-b+1)$. For the special case $\beta=0, \alpha_{1}=....=\alpha_{r}=\gamma_{_{\ell}}+\frac{1}{2}, m_{1}=....=m_{r}=n, x=qt, t_{1}=....=t_{r}=\frac{1}{q}, \mu=q \gamma_{_{\ell}}+\frac{1}{2}, r=2q$, this general relation yields the following results
\begin{equation}
(qt)^{q \gamma_{_{\ell}}+\frac{1}{2}}\left[L_{n}^{\gamma_{_{\ell}}+\frac{1}{2}}(t) \right]^{2q}=\sum_{k=0}^{\infty}\Theta_{k}\left(q \gamma_{_{\ell}}+\frac{1}{2},0, 2q, {n}, {\gamma_{_{\ell}}+\frac{1}{2}},{\frac{1}{q}} \right) L_{k}^{(0)}(qt), 
\label{ps33} 
\end{equation}
where 
\begin{equation}
\Theta_{k}\left(q \gamma_{_{\ell}}+\frac{1}{2},0, 2q, {n}, {\gamma_{_{\ell}}+\frac{1}{2}},{\frac{1}{q}} \right)= \begin{array}{l}
\Gamma(q \gamma_{_{\ell}}+\frac{3}{2})
\left(\begin{array}{c}
n+\gamma_{_{\ell}}+\frac{1}{2}\\
n
\end{array} \right)^{2q} \times\\
F_{A}^{(2q+1)} \left[ q \gamma_{_{\ell}}+\frac{3}{2},-n,...,-n,-k;\gamma_{_{\ell}}+\frac{3}{2},...,\gamma_{_{\ell}}+\frac{3}{2},1;\frac{1}{q},...,\frac{1}{q},1 \right]. 

\end{array}
 \label{ps34}
\end{equation}
By considering $F_{q}[\rho]$, equation (\ref{ps33}) and the orthogonality relation of the Laguerre polynomials, one finds that only the term with $k=0$ gives a non-vanishing contribution to the summation in equation (\ref{ps33}), so that

\begin{equation}
F_{q}[\rho]=\frac{1}{q^{q\gamma_{_{\ell}}+\frac{3}{2}}} \, \Theta_{0}\left(q \gamma_{_{\ell}}+\frac{1}{2},0, 2q, {n}, {\gamma_{_{\ell}}+\frac{1}{2}},{\frac{1}{q}} \right),
 \label{ps35}
\end{equation}
where
\begin{equation}
\Theta_{0}\left(q \gamma_{_{\ell}}+\frac{1}{2},0, 2q, {n}, {\gamma_{_{\ell}}+\frac{1}{2}},{\frac{1}{q}} \right)= \begin{array}{l}
\Gamma(q \gamma_{_{\ell}}+\frac{3}{2})
\left(\begin{array}{c}
n+\gamma_{_{\ell}}+\frac{1}{2}\\
n
\end{array} \right)^{2q} \times\\
F_{A}^{(2q+1)} \left[ q \gamma_{_{\ell}}+\frac{3}{2},-n,...,-n,0;\gamma_{_{\ell}}+\frac{3}{2},...,\gamma_{_{\ell}}+\frac{3}{2},1;\frac{1}{q},...,\frac{1}{q},1 \right]. 

\end{array}
 \label{ps36}
\end{equation}
Hence
\begin{equation}
W_{q}[\rho]=\frac{2 \pi N_{n,\ell}^{2q}}{(2 \lambda q)^{q \gamma_{_{\ell}}+\frac{3}{2}}} \,  \Theta_{0}\left(q \gamma_{_{\ell}}+\frac{1}{2},0, 2q, {n}, {\gamma_{_{\ell}}+\frac{1}{2}},{\frac{1}{q}} \right),
\label{ps37}
\end{equation}
so that the Renyi entropy for the pseudoharmonic potential in the position space is given by
\begin{equation}
 R_{q}[\rho]=\frac{1}{1-q}\log \left[\frac{2 \pi N_{n,\ell}^{2q}}{(2 \lambda q)^{q \gamma_{_{\ell}}+\frac{3}{2}}} \,  \Theta_{0}\left(q \gamma_{_{\ell}}+\frac{1}{2},0, 2q, {n}, {\gamma_{_{\ell}}+\frac{1}{2}},{\frac{1}{q}} \right) \right] .
 \label{ps38} 
 \end{equation} 
 
\begin{figure}[h]
\caption{The variation of the Renyi entropy in the position space with $q$ and $\ell$ when $n=0$}
\centering
\includegraphics[scale=0.7]{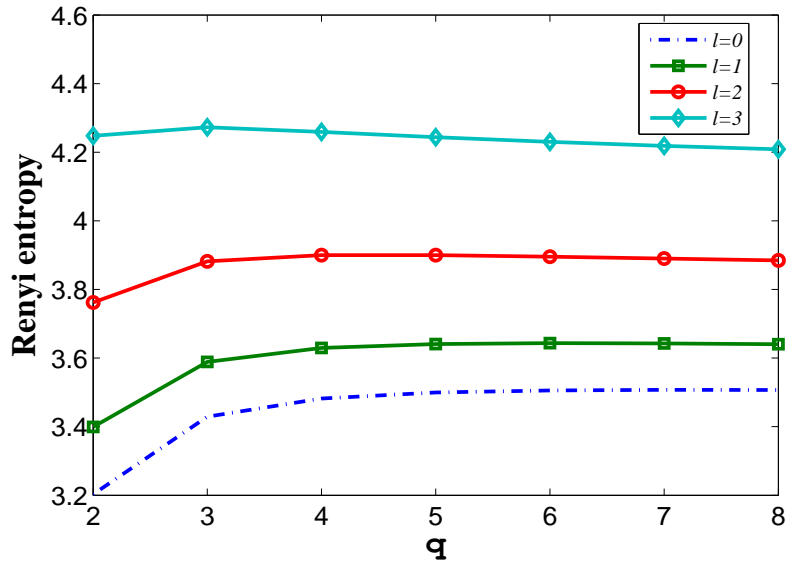}
\label{fig3}
\end{figure}

Note that the Lauricella function of type A is defined as \cite{Deh92, Tua92, Sai95}
\begin{equation}
F_{A}^{(s)}\left(\begin{array}{c}
a;b_{1},...,b_{s}\\
c_{1},...c_{s}
\end{array};x_{1},...,x_{s} \right)=\sum_{j_{1},....,j_{s}=0}^{\infty}\frac{(a)_{j_{1}+....+j_{s}}(b_{1})_{j_{1}}....(b_{s})_{j_{s}}}{(c_{1})_{j_{1}}....(c_{s})_{j_{s}}}\frac{x_{1}^{j_{1}}....x_{s}^{j_{s}}}{j_{1}!....j_{s}!}.
\label{ps39} 
\end{equation}
Table \ref{tab6} contains the numerical values of the Renyi entropy in the position space for some diatomic molecules with $q=2$ and $\ell=0$. Table \ref{tab7} shows the variations with the parameter $q$ of the Renyi entropy in the position space for $NiC$ with $\ell=0$. We observe from Table \ref{tab6} that the Renyi entropy $R_{2}[\rho]$ increases with $n$. This is not the case when $q>2$. Figure \ref{fig3} shows the variation of the position space Renyi entropy with the parameter $q$ and $n=0$.

\subsection{Renyi entropy in the momentum space}
The Renyi entropy in the momentum space is written as

\begin{equation}
R_{q}[\gamma]=\frac{1}{1-q}\log W_{q}[\gamma],
\label{ps40}
\end{equation}
where
\begin{equation}
W_{q}[\gamma]=4 \pi \int _{0}^{\infty} p^{2} \gamma(p)^{q}dp
\label{ps41}
\end{equation}
and $\gamma(p)$ is as given in equation (\ref{ps17}). If we let $\frac{p^{2}}{2\lambda}=x$, then we have
\begin{equation}
W_{q}[\gamma]=2^{q+1} \pi (2\lambda)^{q \gamma_{_{\ell}}+\frac{3}{2}}\left((2\lambda^{2})^{\frac{-q\gamma_{_{\ell}}}{2}-\frac{3}{4}} \right) ^{q}\left[\frac{n!}{\Gamma\left(n+\gamma_{_{\ell}}+\frac{3}{2} \right) } \right] ^{q} \int_{0}^{\infty} x^{q \gamma_{_{\ell}}+\frac{1}{2}}\, e^{-4qx}\left[L_{n}^{\gamma_{_{\ell}}+\frac{1}{2}}(x) \right] ^{2q}dx,
\label{ps42}
\end{equation}
which can be written as
\begin{equation}
W_{q}[\gamma]=2^{q+1} \pi (2\lambda)^{q \gamma_{_{\ell}}+\frac{3}{2}}\left((2\lambda^{2})^{\frac{-q\gamma_{_{\ell}}}{2}-\frac{3}{4}} \right) ^{q}\times  H_{q}[\gamma],
\label{ps43}
\end{equation}
where
\begin{equation}
H_{q}[\gamma]=\left[\frac{n!}{\Gamma\left(n+\gamma_{_{\ell}}+\frac{3}{2} \right) } \right] ^{q} \int_{0}^{\infty} x^{q \gamma_{_{\ell}}+\frac{1}{2}}\, e^{-4qx}\left[L_{n}^{\gamma_{_{\ell}}+\frac{1}{2}}(x) \right] ^{2q}dx.
 \label{ps44}
\end{equation}
$H_{q}[\gamma]$ will be evaluated by using the multivariate Bell polynomials of Combinatorics. We follow the approach similar to Ref. \cite{San11}, whereby, we consider the explicit expression of the Laguerre polynomials given by
\begin{equation}
\tilde{L}_{n}^{\gamma_{_{\ell}}+\frac{1}{2}}(x)=\sum _{k=0}^{n}c_{k}^{(n,\gamma_{_{\ell}}+\frac{1}{2})}x^{k}
\label{ps45}
\end{equation}
with
\begin{equation}
c_{k}^{(n,\gamma_{_{\ell}}+\frac{1}{2})}=\sqrt{\frac{\Gamma(n+\gamma_{_{\ell}}+\frac{3}{2})}{n!}}\frac{(-1)^{k}}{\Gamma(\gamma_{_{\ell}}+k+\frac{3}{2})}\left(\begin{array}{c}
n\\
k
\end{array} \right). 
\label{ps46}
\end{equation}

The Laguerre polynomials can be written in terms of the multivariate Bell polynomials of Combinatorics as \cite{San11, San10}
\begin{equation}
\left[ \tilde{L}_{n}^{\gamma_{_{\ell}}+\frac{1}{2}}(x)\right] ^{2q}=\sum_{k=0}^{2nq}\frac{(2q)!}{(k+2q)!}B_{k+2q,2q}\left(c_{0}^{(n,\gamma_{_{\ell}}+\frac{1}{2})},2! c_{1}^{(n,\gamma_{_{\ell}}+\frac{1}{2})},...,(k+1)!c_{k}^{(n,\gamma_{_{\ell}}+\frac{1}{2})} \right)x^{k}, 
\label{ps47}
\end{equation}
with $c_{i}^{(n,\gamma_{_{\ell}}+\frac{1}{2})}=0$ for $i>n$ and the remaining coefficients are given by equation (\ref{ps46}). The Bell polynomials are given by \cite{San11, San10, Bou09, Ber05}
\begin{equation}
 B_{m,t}(a_{1},a_{2},...,a_{m-t+1})=\sum_{\hat{\pi}(m,t)}\frac{m!}{j_{1}!j_{2}!...j_{m-t+1}!}\left(\frac{a_{1}}{1!} \right) ^{j_{1}}\left(\frac{a_{2}}{2!} \right) ^{j_{2}}...\left(\frac{a_{m-t+1}}{(m-t+1)!} \right) ^{j_{m-t+1}},
 \label{ps48}
 \end{equation} 
where the sum runs over all partitions $\hat{\pi}(m,t)$ such that
\[j_{1}+j_{2}+...+j_{m-t+1}=t, \, \mbox{and} \, j_{1}+2j_{2}+...+(m-t+1)j_{m-t+1}=m. \]  

Therefore,
\begin{equation}
H_{q}[\gamma]=\sum_{k=0}^{2nq}\frac{(2q)!}{(k+2q)!}B_{k+2q,2q}\left(c_{0}^{(n,\gamma_{_{\ell}}+\frac{1}{2})},2! c_{1}^{(n,\gamma_{_{\ell}}+\frac{1}{2})},...,(k+1)!c_{k}^{(n,\gamma_{_{\ell}}+\frac{1}{2})} \right)\, \int_{0}^{\infty}x^{q \gamma_{_{\ell}}+k+\frac{1}{2}}\, e^{-4qx}dx.
\label{ps49}
\end{equation}
On evaluating the integral in equation (\ref{ps49}), we obtain
\begin{equation}
H_{q}[\gamma]=\sum_{k=0}^{2nq} \frac{\Gamma(q\gamma_{_{\ell}}+k+\frac{3}{2})}{2^{2q\gamma_{_{\ell}}+2k+3} q^{q\gamma_{_{\ell}}+k+\frac{3}{2}}} \frac{(2q)!}{(k+2q)!}B_{k+2q,2q}\left(c_{0}^{(n,\gamma_{_{\ell}}+\frac{1}{2})},2! c_{1}^{(n,\gamma_{_{\ell}}+\frac{1}{2})},...,(k+1)!c_{k}^{(n,\gamma_{_{\ell}}+\frac{1}{2})} \right).
\label{ps50}
\end{equation}
The only non-vanishing terms correspond to those with $j_{i+1}=0$ so that $\left(c_{i}^{n,\gamma_{_{\ell}}+\frac{1}{2}} \right)^{j_{i}+1}=1 $ for every $i>n$. Hence, substituting equation (\ref{ps50}) into equation (\ref{ps43}) and then substituting equation (\ref{ps43}) into equation (\ref{ps40}) gives the Renyi entropy in the momentum space for the pseudoharmonic potential which can be written as
\begin{equation}
\begin{array}{l}
R_{q}[\rho]=\frac{1}{1-q}\log\left[  2^{q+1} \pi (2\lambda)^{q \gamma_{_{\ell}}+\frac{3}{2}}\left((2\lambda^{2})^{\frac{-q\gamma_{_{\ell}}}{2}-\frac{3}{4}} \right) ^{q}\times \right. \\ 
\left.\sum_{k=0}^{2nq} \frac{\Gamma(q\gamma_{_{\ell}}+k+\frac{3}{2})}{2^{2q\gamma_{_{\ell}}+2k+3} q^{q\gamma_{_{\ell}}+k+\frac{3}{2}}} \frac{(2q)!}{(k+2q)!}B_{k+2q,2q}\left(c_{0}^{(n,\gamma_{_{\ell}}+\frac{1}{2})},2! c_{1}^{(n,\gamma_{_{\ell}}+\frac{1}{2})},...,(k+1)!c_{k}^{(n,\gamma_{_{\ell}}+\frac{1}{2})} \right)\right]
\end{array}. 
\label{ps51}
\end{equation}

The numerical values of the Renyi entropy for the pseudoharmonic potential in the momentum space for some diatomic molecules when $q=2$ are shown in the Table \ref{tab8}. It is observed that the Renyi entropy for the pseudoharmonic potential in the momentum space, unlike in the position space, decreases with increase in $n$.

\subsection{Tsallis entropy for the pseudoharmonic potential}
\label{sec5}
We shall obtain, in this Subsection, the Tsallis entropy for the pseudoharmonic potential in the position and momentum spaces.
\subsubsection{Tsallis entropy in the position space}

The Tsallis entropy in the position space is written as
\begin{equation}
T_{q}[\rho]=\frac{1}{q-1}\left[1-W_{q}[\rho] \right],
\label{ps52} 
\end{equation}
where $W_{q}[\rho]$ is as given in equation (\ref{ps30}). Using equation (\ref{ps37}), the Tsallis entropy in the position space is then obtained as
\begin{equation}
T_{q}[\rho]=\frac{1}{q-1}\left[1-\frac{2 \pi N_{n,\ell}^{2q}}{(2 \lambda q)^{q \gamma_{_{\ell}}+\frac{3}{2}}} \,  \Theta_{0}\left(q \gamma_{_{\ell}}+\frac{1}{2},0, 2q, {n}, {\gamma_{_{\ell}}+\frac{1}{2}},{\frac{1}{q}} \right) \right].
 \label{ps53}
\end{equation}

The numerical values of the Tsallis entropy in the position space is shown in Table \ref{tab9} with $q=2$ and $\ell=0$. We observe that (See Table \ref{tab10}) the value of the Tsallis entropy for the pseudoharmonic potential in the position space tends to a particular value as $\ell$ increases for each $q$ and it is independent of $\ell$ when $q>>1$.

\subsubsection{Tsallis entropy in the momentum space}
The Tsallis entropy in the momentum space can be written as
\begin{equation}
T_{m}[\gamma]=\frac{1}{m-1}\left[ 1-W_{m}[\gamma]\right] ,
 \label{ps54}
\end{equation}
where
\begin{equation}
W_{m}[\gamma]=4 \pi \int_{0}^{\infty} p^{2} \gamma(p)^{m}dp=2^{m+1} \pi (2\lambda)^{m \gamma_{_{\ell}}+\frac{3}{2}}\left((2\lambda^{2})^{\frac{-m\gamma_{_{\ell}}}{2}-\frac{3}{4}} \right) ^{m}\times  H_{m}[\gamma],
 \label{ps55}
\end{equation}
\begin{equation}
H_{m}[\gamma]=\sum_{k=0}^{2nm} \frac{\Gamma(m\gamma_{_{\ell}}+k+\frac{3}{2})}{2^{2m\gamma_{_{\ell}}+2k+3} m^{m\gamma_{_{\ell}}+k+\frac{3}{2}}} \frac{(2m)!}{(k+2m)!}B_{k+2m,2m}\left(c_{0}^{(n,\gamma_{_{\ell}}+\frac{1}{2})},2! c_{1}^{(n,\gamma_{_{\ell}}+\frac{1}{2})},...,(k+1)!c_{k}^{(n,\gamma_{_{\ell}}+\frac{1}{2})} \right),
 \label{ps56}
\end{equation}
and
\[\frac{1}{q}+\frac{1}{m}=2.\]
Table \ref{tab11} shows the numerical values of the Tsallis entropy for the pseudoharmonic potential in the momentum space for $m=\frac{2}{3}$ and $\ell=0$.

\subsection{Onicescu Information Energy for the pseudoharmonic potential}
\label{sec6}
\subsubsection{Onicescu information energy for the pseudoharmonic potential in the position space}
The Onicescu information energy in the position space is written from equation (\ref{ps5}) as
\begin{equation}
E[\rho]=4 \pi \int_{0}^{\infty} \rho^{2}(r)dr,
\label{ps57}
\end{equation}
which is equivalent to $W_{2}[\rho]$ in equations (\ref{ps29}) and (\ref{ps30}) when $q=2$. From equation (\ref{ps37}), $W_{2}[\rho]$ is obtained as
\begin{equation}
 W_{2}[\rho]=\frac{2 \pi N_{n,\ell}^{4}}{(4 \lambda)^{2 \gamma_{_{\ell}}+\frac{3}{2}}} \Theta_{0}\left(2 \gamma_{_{\ell}}+\frac{1}{2},0, 4, {n}, {\gamma_{_{\ell}}+\frac{1}{2}},{\frac{1}{2}} \right).
 \label{ps58}
 \end{equation} 
Therefore, the Onicescu information energy for the pseudoharmonic potential in the position space is given as
\begin{equation}
E[\rho]=\frac{2 \pi N_{n,\ell}^{4}}{(4 \lambda)^{2 \gamma_{_{\ell}}+\frac{3}{2}}} \Theta_{0}\left(2 \gamma_{_{\ell}}+\frac{1}{2},0, 4, {n}, {\gamma_{_{\ell}}+\frac{1}{2}},{\frac{1}{2}} \right).
 \label{ps59}
\end{equation}

\subsubsection{Onicescu information energy for the pseudoharmonic potential in the momentum space}
The Onicescu information energy in the momentum space is written from equation (\ref{ps6}) as
\begin{equation}
E[\gamma]=4 \pi \int_{0}^{\infty} \gamma^{2}(p)dp,
 \label{ps60}
\end{equation}
which is equivalent to $W_{2}[\gamma]$ in equations (\ref{ps41}), (\ref{ps42}) and (\ref{ps43}) when $q=2$. $W_{2}[\gamma]$ is obtained from equation (\ref{ps43}) as
\begin{equation}
W_{2}[\gamma]=2^{3} \pi (2\lambda)^{2 \gamma_{_{\ell}}+\frac{3}{2}}\left((2\lambda^{2})^{\frac{-2\gamma_{_{\ell}}}{2}-\frac{3}{4}} \right) ^{2}\times  H_{2}[\gamma],
 \label{ps61}
\end{equation}
where
\begin{equation}
H_{2}[\gamma]=\sum_{k=0}^{4n} \frac{\Gamma(2\gamma_{_{\ell}}+k+\frac{3}{2})}{2^{4\gamma_{_{\ell}}+2k+3} 2^{2\gamma_{_{\ell}}+k+\frac{3}{2}}} \frac{(4)!}{(k+4)!}B_{k+4,4}\left(c_{0}^{(n,\gamma_{_{\ell}}+\frac{1}{2})},2! c_{1}^{(n,\gamma_{_{\ell}}+\frac{1}{2})},...,(k+1)!c_{k}^{(n,\gamma_{_{\ell}}+\frac{1}{2})} \right).
 \label{ps62}
\end{equation}

The numerical values of the Onicescu information energy for the pseudoharmonic potential in the position and momentum spaces are shown in Table \ref{tab12}. The Onicescu information energy for the pseudoharmonic potential decreases with increase in $n$ for the position space and increases with increasing $n$ for the momentum space.

\section{Information-theoretic lengths and impetuses}
\label{sec4}
We shall obtain, in this Section, the ratio of the information-theoretic impetuses to lengths for the Fisher information, the Shannon information entropy and the Renyi information entropy for the pseudoharmonic potential.

\subsection{Ratio of the Fisher impetus to length for the pseudoharmonic potential}
The ratio of the Fisher impetus to length is defined as \cite{Jac10}:
\begin{equation}
\frac{F_{I}}{F_{L}}=\frac{\sqrt{I[\rho]}}{\sqrt{I[\gamma]}}
 \label{ps63}
\end{equation}
$I[\rho]$  and $I[\gamma]$ are given by equations (\ref{ps19}) and (\ref{ps21}), respectively. 

\subsection{Ratio of the Shannon impetus to length for the pseudoharmonic potential}
The ratio of the Shannon impetus to length is defined as \cite{Jac10}:
\begin{equation}
\frac{S_{I}}{S_{L}}=\left[e^{S[\gamma]-S[\rho]} \right] ^{\frac{1}{3}}
 \label{ps64}
\end{equation}
$S[\rho]$ and $S[\gamma]$ are given by equations (\ref{ps23}) and (\ref{ps27}), respectively.
\subsection{The ratio of the Renyi impetus to length for the pseudoharmonic potential}
\begin{figure}[h]
\caption{The graph of the ratio of fisher impetus to length for the pseudoharmonic potential against n using $NO^{+}$}
\centering
\includegraphics[scale=0.8]{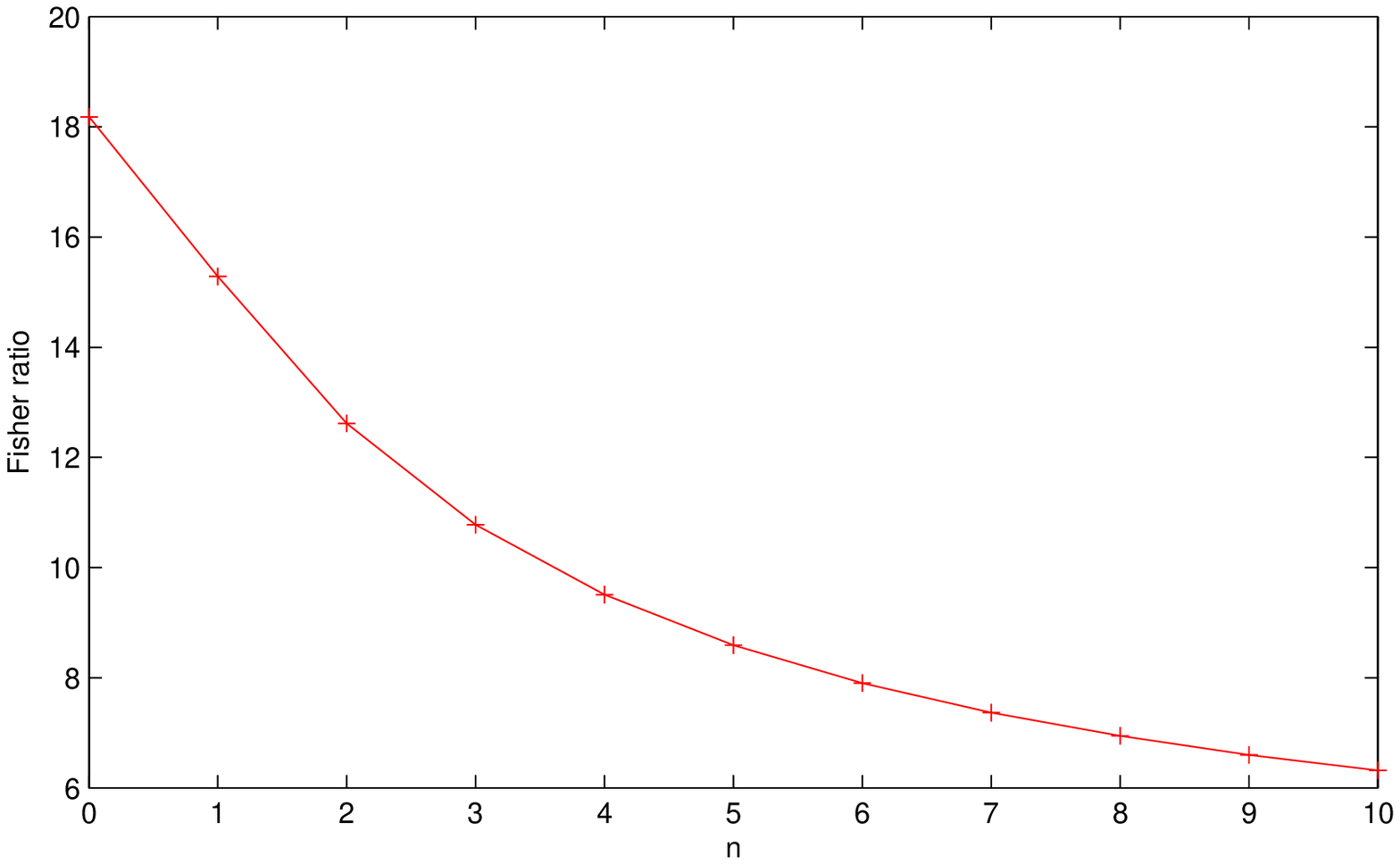}
\label{fig4}
\end{figure}

\begin{figure}[h]
\caption{The graph of the ratio of shannon impetus to length for the pseudoharmonic potential against n using $NO^{+}$}
\centering
\includegraphics[scale=0.8]{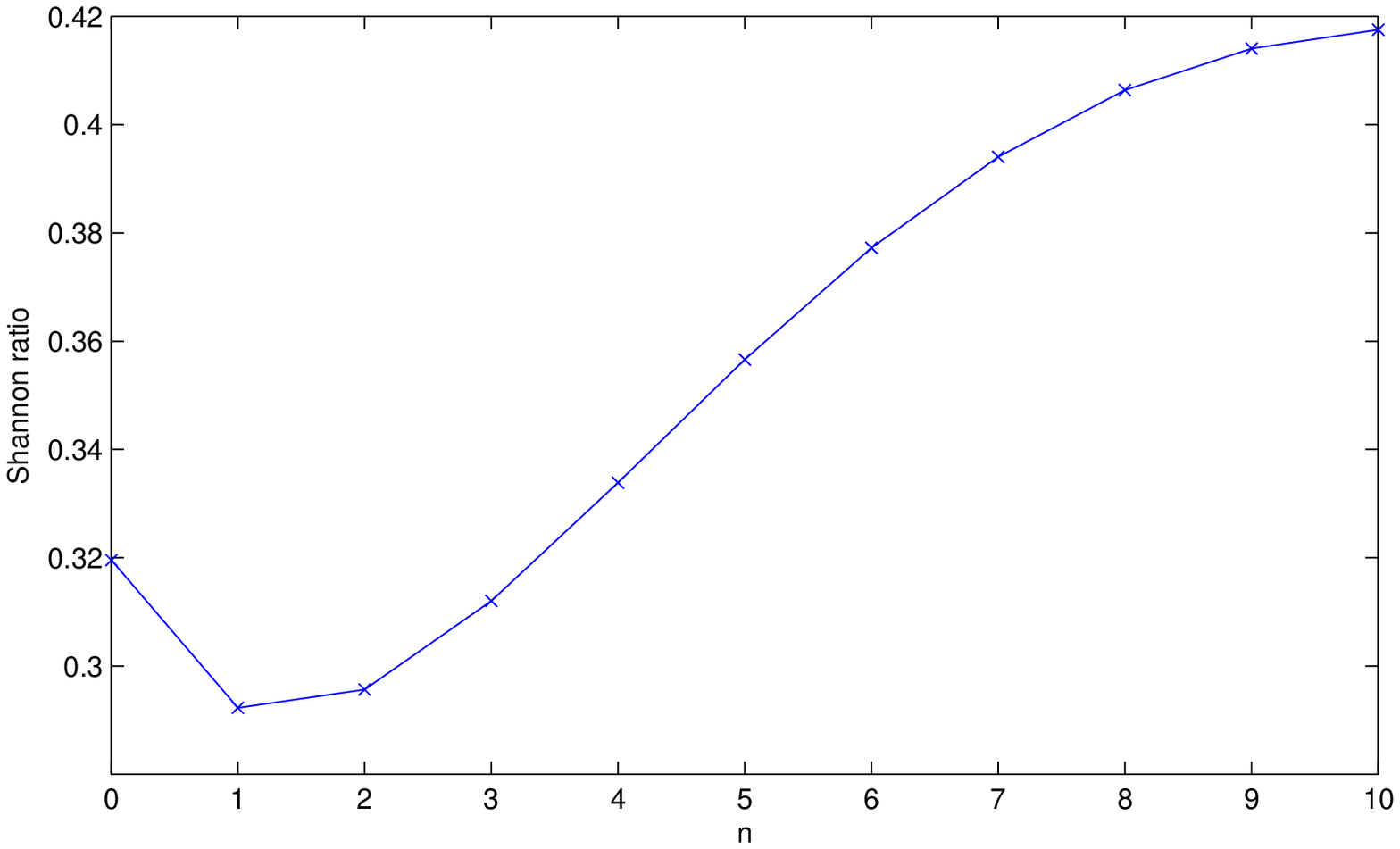}
\label{fig5}
\end{figure}

\begin{figure}[h]
\caption{The graph of the ratio of Renyi impetus to length for the pseudoharmonic potential against n using $NO^{+}$}
\centering
\includegraphics[scale=0.8]{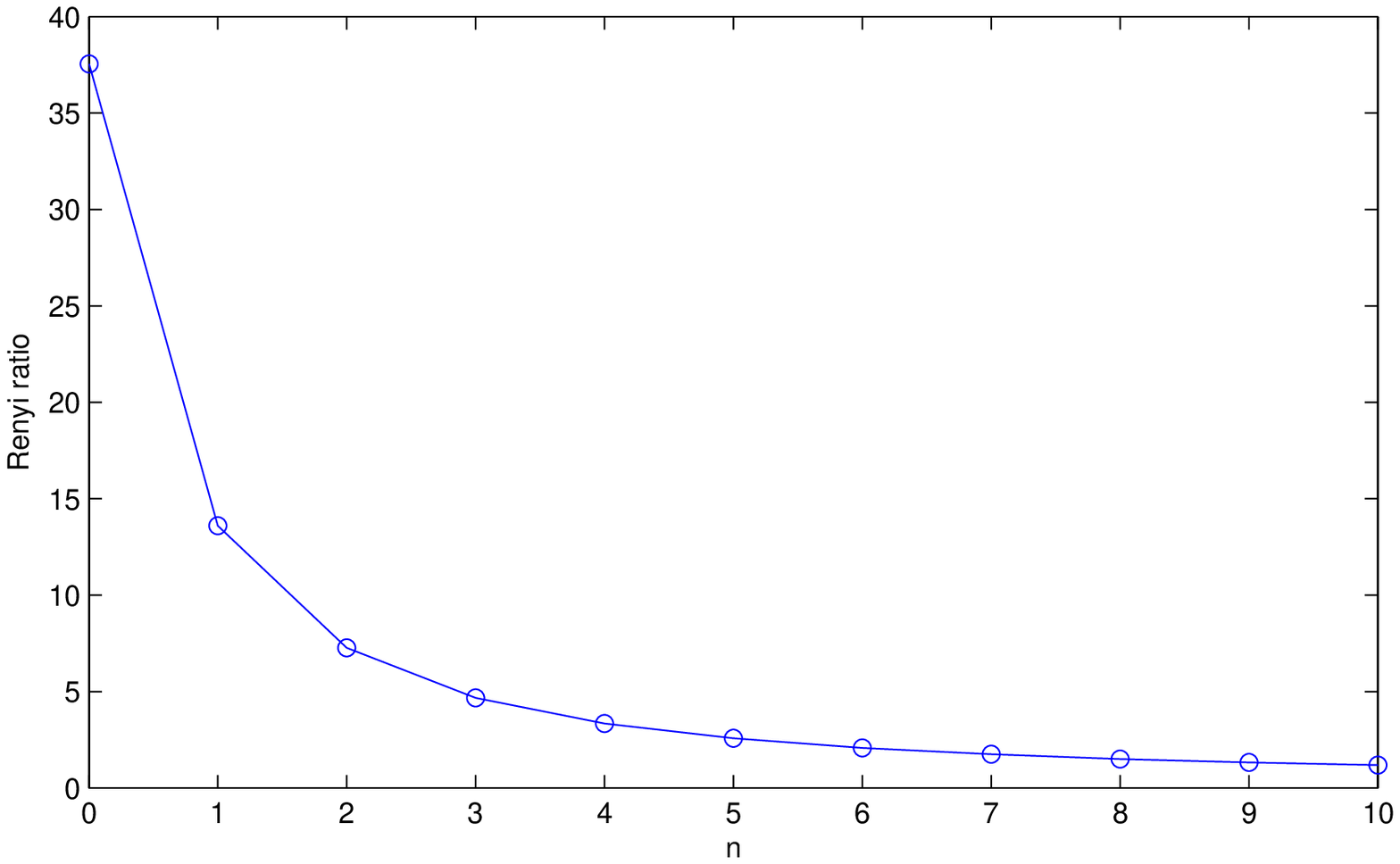}
\label{fig6}
\end{figure}

The ratio of the $q$th-order Renyi impetus length is defined as \cite{Jac10}:
\begin{equation}
\frac{R_{I}}{R_{L}}=\left[e^{R_{q}[\gamma]-R_{q}[\rho]} \right] ^{\frac{1}{3}}
 \label{ps65}
\end{equation}
$R_{q}[\rho]$ and $R_{q}[\gamma]$ are given by equations (\ref{ps38}) and (\ref{ps51}), respectively.
The graphs of the ratio of the information impetuses to lengths of Fisher information $\left(\frac{F_{I}}{F_{L}} \right)$, Shannon entropy $\left(\frac{S_{I}}{S_{L}} \right)$ and Renyi entropy $\left(\frac{R_{I}}{R_{L}} \right)$ for the pseudoharmonic potential when $q=2$ are shown in the Figures \ref{fig4}-\ref{fig6}. From Figure \ref{fig4}, it is observed that the Fisher ratio $\left(\frac{F_{I}}{F_{L}} \right)$ decreases exponentially with increasing $n$. It is observed from Figure \ref{fig5} that the Shannon ratio $\left(\frac{S_{I}}{S_{L}} \right)$ starts to decrease from $n=0$ to $n=1$ but increases with $n$ afterwards. Also, from Figure \ref{fig6}, the Renyi ratio $\left(\frac{R_{I}}{R_{L}} \right)$ for $q=2$ decreases exponentially with increasing $n$.

\section{Conclusion}
\label{sec5}
We have studied, both analytically and numerically, the information-theoretic measures for the pseudoharmonic potential using Fisher information, Shannon entropy, Renyi entropy, Tsallis entropy and Onicescu information energy in both the position and momentum spaces. By using the values of $D_{e}$ and $r_{e}$ of some diatomic molecules, we have obtained the numerical values of the information-theoretic measures for the pseudoharmonic potential. By using the linearization formula of Srivastava-Niukkanen, we have obtained the Renyi and Tsallis entropies in the position space in terms of the Lauricella hypergeometric function. 

To obtain the Renyi and Tsallis entropies in the momentum space, we have employed the use of the multivariate Bell polynomials of Combinatorics. We observed that the Fisher information increases with $n$ in both the position and momentum spaces, but decreases with increasing $\ell$ for all the diatomic molecules considered. 
The Shannon entropy also increases with $n$ in the position space and decreases with increasing $\ell$.

The Renyi entropy increases with the parameter $q$ but the Tsallis entropy decreases with increase in the parameter $q$. We also observed that as $n$ increases, the Tsallis entropy tends to a particular value for each $q$. For large value of $q$, the Tsallis entropy is constant with change in $n$. The analytical and numerical values of the Onicescu information energy have also been obtained and we finally obtained the ratio of the information-theoretic impetuses to lengths of Fisher, Shannon and Renyi.

\begin{center}
\begin{table}
\caption{Spectroscopic constants of some diatomic molecules}
{\begin{tabular}{lcc}\hline\hline  \\
molecular state & $D_{e}(eV)$ & $r_{e}(A^{0})$ \\ 
\hline\hline \\
$Na_{2}: X^{1}\Sigma_{g}^{+}$ & 0.746707167 & 3.079 \\[1ex] 
 
$Cl_{2}:X^{1}\Sigma_{g}^{+}$ & 2.513903386 & 1.987 \\[1ex] 
 
$O_{2}^{+}:X^{2}I_{g}$ & 6.780447246 & 1.116 \\[1ex]
 
$N_{2}^{+}:X^{2}\Sigma_{g}^{+}$ & 8.848131541 & 1.116 \\[1ex] 
 
$NO^{+}:X^{1}\Sigma^{+}$ & 10.99665353 & 1.063 \\[1ex]
\hline\hline 

\end{tabular} 
\label{tab1}}
\end{table}
\end{center}

\begin{center}
\begin{table}
\caption{Fisher information entropy for the pseudoharmonic potential for some diatomic molecules with $\ell=0$ in the position space. The unit of the square of $I[\rho]$ is $eV/\left( A^{o}\right) ^{2}$.}
{\begin{tabular}{cccccc}
\hline\hline  \\
$n$ & $I[\rho](Na_{2})$ & $I[\rho](Cl_{2})$ & $I[\rho](O_{2}^{+})$ & $I[\rho](N_{2}^{+})$ & $I[\rho](NO^{+})$ \\ 
\hline\hline \\
0 & 4.03449 & 8.95359 & 29.6189 & 27.4679 & 29.0007 \\[1ex]

1 & 11.6049 & 25.9150 & 85.4832 & 79.6411 & 84.2456 \\ [1ex]

2 & 19.1753 & 42.8764 & 141.348 & 131.814 & 139.490 \\ [1ex]
 
3 & 26.7456 & 59.8378 & 197.212 & 183.987 & 194.735 \\ [1ex]
 
4 & 34.3160 & 76.7992 & 253.076 & 236.161 & 249.980 \\ [1ex]
 
5 & 41.8864 & 93.7605 & 308.941 & 288.334 & 305.225 \\ [1ex]
 
6 & 49.4568 & 110.722 & 364.805 & 340.507 & 360.470 \\ [1ex]
 
7 & 57.0272 & 127.683 & 420.669 & 392.680 & 415.715 \\ [1ex]
 
8 & 64.5976 & 144.645 & 476.534 & 444.853 & 470.960 \\ [1ex]
 
9 & 72.1680 & 161.606 & 532.398 & 497.026 & 526.204 \\ [1ex]
 
10 & 79.7383 & 178.567 & 588.262 & 549.199 & 581.449 \\ [1ex]
\hline\hline 
\end{tabular} 
\label{tab2}}
\end{table}
\end{center}

\begin{center}
\begin{table}
\caption{Fisher information entropy for the pseudoharmonic potential for some diatomic molecules with $\ell=0$ in the momentum space. The unit of the square of $I[\gamma]$ is $\left( A^{o}\right) ^{2}/eV$.}
{\begin{tabular}{cccccc}
\hline\hline \\
$n$ & $I[\gamma](Na_{2})$ & $I[\gamma](Cl_{2})$ & $I[\gamma](O_{2}^{+})$ & $I[\gamma](N_{2}^{+})$ & $I[\gamma](NO^{+})$ \\ 
\hline\hline \\
0 & 3.68815 & 0.62866 & 0.30859 & 0.14658 & 0.09247 \\[1ex]

1 & 11.9955 & 2.28850 & 1.06359 & 0.55325 & 0.36408 \\ [1ex]

2 & 24.5468 & 5.10871 & 2.27526 & 1.27083 & 0.86499 \\ [1ex]
 
3 & 40.6289 & 9.05928 & 3.89987 & 2.30608 & 1.61329 \\ [1ex]
 
4 & 59.5473 & 14.0403 & 5.87902 & 3.64230 & 2.60660 \\ [1ex]
 
5 & 80.7177 & 19.9301 & 8.15580 & 5.25260 & 3.83137 \\ [1ex]
 
6 & 103.670 & 26.6080 & 10.6799 & 7.10709 & 5.26891 \\ [1ex]
 
7 & 128.032 & 33.9629 & 13.4085 & 9.17643 & 6.89883 \\ [1ex]
 
8 & 153.505 & 41.8964 & 16.3058 & 11.4334 & 8.70091 \\ [1ex]
 
9 & 179.852 & 50.3228 & 19.3421 & 13.8535 & 10.6560 \\ [1ex]
 
10 & 206.884 & 59.1681 & 22.4924 & 16.4150 & 12.7465 \\ [1ex]
\hline\hline 
\end{tabular} 
\label{tab3}}
\end{table}
\end{center}

\begin{center}
\begin{table}
\caption{The Shannon information entropy for the pseudoharmonic potential for some diatomic molecules with $\ell=0$ in the position space.}
{\begin{tabular}{cccccc}
\hline\hline \\
$n$ & $S[\rho](Na_{2})$ & $S[\rho](Cl_{2})$ & $S[\rho](O_{2}^{+})$ & $S[\rho](N_{2}^{+})$ & $S[\rho](NO^{+})$ \\ 
\hline\hline \\
0 & 12.4333 & 7.58285 & 4.73017 & 4.07721 & 3.56137 \\[1ex]

1 & 13.5554 & 8.41921 & 5.69732 & 4.83464 & 4.23386 \\ [1ex]

2 & 14.3281 & 8.99397 & 6.36283 & 5.35458 & 4.69475 \\ [1ex]
 
3 & 14.9377 & 9.44880 & 6.88877 & 5.76628 & 5.05992 \\ [1ex]
 
4 & 15.4466 & 9.83007 & 7.32886 & 6.11179 & 5.36671 \\ [1ex]
 
5 & 15.8856 & 10.1602 & 7.70925 & 6.41130 & 5.63298 \\ [1ex]
 
6 & 16.2725 & 10.4522 & 8.04514 & 6.67647 & 5.86899 \\ [1ex]
 
7 & 16.6188 & 10.7145 & 8.34633 & 6.91483 & 6.08136 \\ [1ex]
 
8 & 16.9326 & 10.9527 & 8.61959 & 7.13156 & 6.27465 \\ [1ex]
 
9 & 17.2196 & 11.1711 & 8.86983 & 7.33042 & 6.45214 \\ [1ex]
 
10 & 17.4841 & 11.3729 & 9.10074 & 7.51422 & 6.61633 \\ [1ex]
\hline\hline 
\end{tabular} 
\label{tab4}}
\end{table}
\end{center}

\begin{table}
\caption{The Shannon information entropy for the pseudoharmonic potential for some diatomic molecules with $\ell=0$ in the momentum space.}
{\begin{tabular}{cccccc}
\hline\hline \\
$n$ & $S[\gamma](Na_{2})$ & $S[\gamma](Cl_{2})$ & $S[\gamma](O_{2}^{+})$ & $S[\gamma](N_{2}^{+})$ & $S[\gamma](NO^{+})$ \\ 
\hline\hline\\ 
0 & 0.316484 & 0.257021 & 0.438126 & 0.269701 & 0.214067 \\[1ex]

1 & 0.581714 & 0.632060 & 1.066130 & 0.737227 & 0.619039 \\ [1ex]

2 & 0.699930 & 1.006880 & 1.698210 & 1.287190 & 1.130940 \\ [1ex]
 
3 & 0.677853 & 1.314390 & 2.237290 & 1.826600 & 1.666120 \\ [1ex]
 
4 & 0.557843 & 1.534280 & 2.656090 & 2.304790 & 2.169300 \\ [1ex]
 
5 & 0.382078 & 1.671710 & 2.962040 & 2.703120 & 2.612640 \\ [1ex]
 
6 & 0.181821 & 1.741740 & 3.174780 & 3.021470 & 2.987110 \\ [1ex]
 
7 & -0.022832& 1.760900 & 3.315330 & 3.268260 & 3.294360 \\ [1ex]
 
8 & -0.220225 & 1.743700 & 3.401900 & 3.454700 & 3.541080 \\ [1ex]
 
9 & -0.404298 & 1.701700 & 3.448900 & 3.591960 & 3.735720 \\ [1ex]
 
10 & -0.572438 & 1.643560 & 3.467290 & 3.689930 & 3.886720 \\ [1ex]
\hline\hline 
\end{tabular} 
\label{tab5}}
\end{table}

\begin{center}
\begin{table}
\caption{Numerical values of the Renyi entropy for the pseudoharmonic potential in the position space for some diatomic molecules with $q=2$ and $\ell=0$.}
{\begin{tabular}{cccccc}
\hline\hline \\
$n$ & $R_{2}[\rho](Na_{2})$ & $R_{2}[\rho](Cl_{2})$ & $R_{2}[\rho](O_{2}^{+})$ & $R_{2}[\rho](N_{2}^{+})$ & $R_{2}[\rho](NO^{+})$ \\ 
\hline\hline\\ 
0 & 4.13494 & 3.20296 & 1.27924 & 1.60833 & 1.62971 \\[1ex]

1 & 4.47401 & 3.53567 & 1.61492 & 1.93919 & 1.95852 \\ [1ex]

2 & 4.66934 & 3.72714 & 1.80821 & 2.12949 & 2.14751 \\ [1ex]
 
3 & 4.80869 & 3.86388 & 1.94620 & 2.26543 & 2.28254 \\ [1ex]
 
4 & 4.91756 & 3.97088 & 2.05410 & 2.37185 & 2.38829 \\ [1ex]
 
5 & 5.00708 & 4.05900 & 2.14290 & 2.45953 & 2.47546 \\ [1ex]
 
6 & 5.08315 & 4.13400 & 2.21842 & 2.53419 & 2.54972 \\ [1ex]
 
7 & 5.14932 & 4.19932 & 2.28415 & 2.59923 & 2.61444 \\ [1ex]
 
8 & 5.20788 & 4.25719 & 2.34236 & 2.65688 & 2.67182 \\ [1ex]
 
9 & 5.26040 & 4.30915 & 2.39459 & 2.70865 & 2.72337 \\ [1ex]
 
10 & 5.30801 & 4.35629 & 2.44196 & 2.75564 & 2.77018 \\ [1ex]
\hline\hline 
\end{tabular} 
\label{tab6}}
\end{table}

\end{center}

\begin{center}
\begin{table}
\caption{Variations of the Renyi entropy for the pseudoharmonic potential in the the position space with the $q$ and $n$ for $Cl_{2}$ molecule with $\ell=0$.}
\begin{tabular}{cccccc}
\hline\hline \\
$n$ & $q=3$ & $q=4$ & $q=5$ & $q=6$ & $q=7$ \\ 
\hline\hline \\ 
0 & 3.42934 & 3.48219 & 3.49964 & 3.50566 & 3.50715  \\ [1ex]

1 & 3.64938 & 3.61909 & 3.58041 & 3.54859 & 3.52373 \\ [1ex]

2 & 3.72728 & 3.63067 & 3.55724 & 3.50571 & 3.46835 \\ [1ex]

3 & 3.76441 & 3.61976 & 3.52367 & 3.45939 & 3.41377  \\ [1ex]

4 & 3.78507 & 3.60376 & 3.49082 & 3.41689 & 3.36493 \\ [1ex]

5 & 3.79775 & 3.58701 & 3.46053 & 3.37873 & 3.32156\\ [1ex]

6 & 3.80610 & 3.57079 & 3.43292 & 3.34443 & 3.28283 \\ [1ex]

7 & 3.81188 & 3.55549 & 3.40774 & 3.31340 & 3.24792 \\ [1ex]

8 & 3.81605 & 3.54116 & 3.38466 & 3.28514 & 3.21620 \\ [1ex]

9 & 3.81915 & 3.52777 & 3.36341 & 3.25921 & 3.18717 \\ [1ex]

10 & 3.82153 & 3.51526 & 3.34375 & 3.23529 & 3.16041 \\ [1ex]
\hline \hline
\end{tabular} 
\label{tab7}
\end{table}
\end{center}

\begin{center}
\begin{table}
\caption{The numerical values of the Renyi entropy for the pseudoharmonic potential in the momentum space for some diatomic molecules when $q=2$.}
{\begin{tabular}{cccccc}
\hline\hline \\
$n$ & $R_{2}[\gamma](Na_{2})$ & $R_{2}[\gamma](Cl_{2})$ & $R_{2}[\gamma](O_{2}^{+})$ & $R_{2}[\gamma](N_{2}^{+})$ & $R_{2}[\gamma](NO^{+})$ \\ 
\hline\hline\\ 
0 & 5.93170 & 9.08768 & 9.90687 & 11.4410 & 12.3349 \\[1ex]

1 & 3.69398 & 6.61131 & 7.54612 & 8.88826 & 9.69331 \\ [1ex]

2 & 2.38924 & 5.11855 & 6.14517 & 7.33419 & 8.06709 \\ [1ex]
 
3 & 1.51135 & 4.08903 & 5.19016 & 6.25443 & 6.82777 \\ [1ex]
 
4 & 0.87653 & 3.32961 & 4.49229 & 5.45320 & 6.07661 \\ [1ex]
 
5 & 0.39591 & 2.74501 & 3.95927 & 4.83330 & 5.41436 \\ [1ex]
 
6 & 0.01995 & 2.28109 & 3.53914 & 4.33926 & 4.88398 \\ [1ex]
 
7 & -0.28152 & 1.90434 & 3.19998 & 3.93653 & 4.44978 \\ [1ex]
 
8 & -0.52805 & 1.59272 & 2.92095 & 3.60229 & 4.08805 \\ [1ex]
 
9 & -0.73290 & 1.33107 & 2.68780 & 3.32079 & 3.78235 \\ [1ex]
 
10 & -0.90541 & 1.10862 & 2.49045 & 3.08079 & 3.52089 \\ [1ex]
\hline\hline 
\end{tabular} 
\label{tab8}}
\end{table}
\end{center}

\begin{center}

\begin{table}
\caption{The numerical values of the Tsallis entropy for the pseudoharmonic potential in the position space with $q=2$ and $\ell=0$.}
{\begin{tabular}{cccccc}
\hline\hline \\
$n$ & $T_{2}[\rho](Na_{2})$ & $T_{2}[\rho](Cl_{2})$ & $T_{2}[\rho](O_{2}^{+})$ & $T_{2}[\rho](N_{2}^{+})$ & $T_{2}[\rho](NO^{+})$ \\ 
\hline\hline\\ 
0 & 0.983996 & 0.959358 & 0.721752 & 0.799777 & 0.804013 \\[1ex]

1 & 0.988599 & 0.970861 & 0.801094 & 0.856180 & 0.858933 \\ [1ex]

2 & 0.990622 & 0.975938 & 0.836053 & 0.881102 & 0.883226 \\ [1ex]
 
3 & 0.991841 & 0.979013 & 0.857184 & 0.896215 & 0.897975 \\ [1ex]
 
4 & 0.992683 & 0.981143 & 0.871792 & 0.906692 & 0.908214 \\ [1ex]
 
5 & 0.993310 & 0.982734 & 0.882685 & 0.914525 & 0.915876 \\ [1ex]
 
6 & 0.993800 & 0.983981 & 0.891219 & 0.920674 & 0.921896 \\ [1ex]
 
7 & 0.994197 & 0.984994 & 0.898139 & 0.925670 & 0.926791 \\ [1ex]
 
8 & 0.994527 & 0.985838 & 0.903899 & 0.929833 & 0.930874 \\ [1ex]
 
9 & 0.994807 & 0.986555 & 0.908790 & 0.933373 & 0.934347 \\ [1ex]
 
10 & 0.995048 & 0.987174 & 0.913010 & 0.936432 & 0.937349 \\ [1ex]
\hline\hline 
\end{tabular} 
\label{tab9}}
\end{table}

\end{center}

\begin{center}
\begin{table}
\caption{The numerical values of the Tsallis entropy for the pseudoharmonic potential in the position space as $q$ increases when $\ell=0$ using $NO^{+}$ molecule.
}
\begin{tabular}{cccccc}
\hline\hline \\
$n$ & $q=3$ & $q=4$ & $q=5$ & $q=6$ & $q=10$ \\ 
\hline\hline \\ 
0 & 0.804013 & 0.485394 & 0.331782 & 0.249808 & 0.111111  \\ [1ex]

1 & 0.835086 & 0.488945 & 0.332277 & 0.249882 & 0.111111 \\ [1ex]

2 & 0.880675 & 0.493431 & 0.332818 & 0.249953 & 0.111111 \\ [1ex]

3 & 0.923289 & 0.496763 & 0.333139 & 0.249986 & 0.111111  \\ [1ex]

4 & 0.954746 & 0.498604 & 0.333272 & 0.249997 & 0.111111 \\ [1ex]

5 & 0.974857 & 0.499451 & 0.333316 & 0.249999 & 0.111111 \\ [1ex]

6 & 0.986594 & 0.499797 & 0.333329 & 0.250000 & 0.111111 \\ [1ex]

7 & 0.993050 & 0.499928 & 0.333332 & 0.250000 & 0.111111 \\ [1ex]

8 & 0.996466 & 0.499975 & 0.333333 & 0.250000 & 0.111111\\ [1ex]

9 & 0.998227 & 0.499991 & 0.333333 & 0.250000 & 0.111111 \\ [1ex]

10 & 0.999119 & 0.499997 & 0.333333 & 0.250000 & 0.111111 \\ [1ex]
\hline \hline
\end{tabular} 
\label{tab10}
\end{table}
\end{center}

\begin{table}
\caption{The numerical values of the Tsallis entropy in the momentum space for $m=\frac{2}{3}$ when $\ell=0$}
{\begin{tabular}{cccccc}
\hline\hline \\
$n$ & $T_{m}[\gamma](Na_{2})$ & $T_{m}[\gamma](Cl_{2})$ & $T_{m}[\gamma](O_{2}^{+})$ & $T_{m}[\gamma](N_{2}^{+})$ & $T_{m}[\gamma](NO^{+})$ \\ 
\hline\hline\\ 
0 & -2.07110 & -1.97990 & -0.83715 & -1.39502 & -1.55384 \\[1ex]

1 & -1.30571 & -0.95051 & 1.14771 & 0.32431 & 0.10281 \\ [1ex]

2 & -0.71571 & -0.01585 & 2.81766 & 1.96173 & 1.76714 \\ [1ex]
 
3 & -0.29502 & 0.75581 & 4.10632 & 3.37149 & 3.26865 \\ [1ex]
 
4 & -0.00556 & 1.36370 & 5.06124 & 4.52435 & 4.54838 \\ [1ex]
 
5 & 0.18981 & 1.83168 & 5.75521 & 5.44275 & 5.60653 \\ [1ex]
 
6 & 0.31935 & 2.18773 & 6.25382 & 6.16456 & 6.46739 \\ [1ex]
 
7 & 0.40295 & 2.45664 & 6.60820 & 6.72764 & 7.16158 \\ [1ex]
 
8 & 0.45433 & 2.65835 & 6.85627 & 7.16459 & 7.71844 \\ [1ex]
 
9 & 0.48290 & 2.80833 & 7.02563 & 7.50188 & 8.16349 \\ [1ex]
 
10 & 0.49521 & 2.91834 & 7.13635 & 7.76048 & 8.51776 \\ [1ex]
\hline\hline 
\end{tabular} 
\label{tab11}}
\end{table}

\begin{table}
\caption{The numerical values of the Onicescu information energy for the pseudoharmonic potential in the position and momentum spaces when $\ell=0$}
{\begin{tabular}{cccccc}
\hline\hline \\
$n$ & $E[\rho](O_{2}^{+})$ & $E[\rho](NO^{+})$ & $E[\gamma](O_{2}^{+})$ &  $E[\gamma](NO^{+})$ \\ 
\hline\hline\\ 
0 & 0.278248 & 0.195987 & 0.000049831 & 0.000004395 \\[1ex]

1 & 0.198906 & 0.141067 & 0.000528157 & 0.000061695 \\ [1ex]

2 & 0.163947 & 0.116774 & 0.002143820 & 0.000313693 \\ [1ex]
 
3 & 0.142816 & 0.102025 & 0.005571110 & 0.000980189 \\ [1ex]
 
4 & 0.128208 & 0.091786 & 0.011195000 & 0.002295950 \\ [1ex]
 
5 & 0.117315 & 0.084124 & 0.019077100 & 0.004452190 \\ [1ex]
 
6 & 0.108781 & 0.078104 & 0.029038400 & 0.007566820 \\ [1ex]
 
7 & 0.101861 & 0.073209 & 0.040763000 & 0.011681100 \\ [1ex]
 
8 & 0.096101 & 0.069126 & 0.053882700 & 0.016771900  \\ [1ex]
 
9 & 0.091210 & 0.065653 & 0.068030600 & 0.022769100  \\ [1ex]
 
10 & 0.086990 & 0.062651 & 0.082872600 & 0.029573100  \\ [1ex]
\hline\hline 
\end{tabular} 
\label{tab12}}
\end{table}

\end{document}